\documentclass[11pt]{article}
\usepackage{amssymb,amsmath,amsfonts}
\usepackage{graphicx}
\usepackage{graphics}
\usepackage{eepic,epsfig}
\usepackage{dcolumn}

\begin{document}

\title{Synchrotron radiation from a charge moving along a helical orbit inside
a dielectric cylinder}
\author{A. A. Saharian$^{1,2}$\thanks{%
Email address: saharyan@server.physdep.r.am} \, and  A. S.
Kotanjyan$^{1}$\\ \\
\textit{$^{1}$Institute of Applied Problems in Physics, 375014 Yerevan, Armenia}\\
\textit{$^{2}$The Abdus Salam International Centre for Theoretical
Physics, 34014 Trieste, Italy }}
\date{\today}
\maketitle

\begin{abstract}
The radiation emitted by a charged particle moving along a helical
orbit inside a dielectric cylinder immersed into a homogeneous
medium is investigated. Expressions are derived for the
electromagnetic potentials, electric and magnetic fields, and for
the spectral-angular distribution of radiation in the exterior
medium. It is shown that under the Cherenkov condition for
dielectric permittivity of the cylinder and the velocity of the
particle image on the cylinder surface, strong narrow peaks are
present in the angular distribution for the number of radiated
quanta. At these peaks the radiated energy exceeds the
corresponding quantity for a homogeneous medium by some orders of
magnitude. The results of numerical calculations for the angular
distribution of radiated quanta are presented and they are
compared with the corresponding quantities for radiation in a
homogeneous medium. The special case of relativistic charged
particle motion along the direction of the cylinder axis with
non-relativistic transverse velocity (helical undulator) is
considered in detail. Various regimes for the undulator parameter
are discussed. It is shown that the presence of the cylinder can
increase essentially the radiation intensity.
\end{abstract}

\bigskip

PACS number(s): 41.60.Ap, 41.60.Bq

\bigskip

\section{Introduction}

\label{sec:oscint}

The unique characteristics of synchrotron radiation, such as high
intensity, high collimation, and the wide spectral range (see
Refs. \cite{Koch,Win80,Soko86,Tern85,Bord99,Hofm04} and references
therein), have resulted in extensive applications of synchrotron
radiation in a wide variety of experiments and in many
disciplines. These applications motivate the importance of
investigations for various mechanisms of controlling the radiation
parameters. From this point of view, it is of interest to consider
the influence of a medium on the spectral and angular
distributions of the synchrotron emission. This study is also
important with respect to some astrophysical problems
\cite{Bord99,Ryb79}. The presence of medium can essentially change
the characteristics of the high energy electromagnetic processes
and gives rise to new types of phenomena. Well-known examples are
Cherenkov, transition, and diffraction radiations. The operation
of a number of devices assigned to production of electromagnetic
radiation is based on the interaction of high-energy electrons
with materials (see, for example, \cite{Rull98}).

The synchrotron radiation from a charged particle circulating in a
homogeneous medium was considered by Tsytovich in Ref.
\cite{Tsytovich} (see also Refs. \cite{Kitao,Zrelov}), where it
had been shown that the interference between the synchrotron and
Cherenkov radiations leads to interesting effects. New interesting
peculiarities arise in the case of inhomogeneous media. In
particular, the interfaces of media can be used to control the
radiation flow emitted by various systems. In a series of papers
started in Refs. \cite{Grigoryan1995,Grig95b} we have considered
the most simple geometries of boundaries, namely, the boundaries
with spherical and cylindrical symmetries. The synchrotron
radiation from a charge rotating around a dielectric ball enclosed
by a homogeneous medium is investigated in Refs.
\cite{Grig95b,Grigoryan1998}. It has been shown that the
interference between the synchrotron and Cherenkov radiations
leads to interesting effects: if for the material of the ball and
the particle velocity the Cherenkov condition is satisfied, there
are strong narrow peaks in the radiation intensity. At these peaks
the radiated energy exceeds the corresponding quantity in the case
of a homogeneous medium by some orders of magnitude. A similar
problem for the case of the cylindrical symmetry has been
considered in Refs.
\cite{Grigoryan1995,Kot2000,Kot2001,Kota02,KotNIMB}. In Ref. \cite
{Grigoryan1995} we have developed a recurrent scheme for
constructing the Green function of the electromagnetic field for a
medium consisting of an arbitrary number of coaxial cylindrical
layers. The investigation of the radiation from a charged particle
circulating around a dielectric cylinder immersed in a homogeneous
medium, has shown that under the Cherenkov condition for the
material of the cylinder and the velocity of the particle, there
are narrow peaks in the angular distribution of the number of
quanta emitted into the exterior space. For some values of the
parameters the density of the number of quanta in these peaks
exceeds the corresponding quantity for the radiation in vacuum by
several orders. The radiation by a longitudinal charged oscillator
moving with a constant drift velocity along the axis of a
dielectric cylinder immersed in a homogeneous medium is
investigated in Refs. \cite{Saha03,Saha04}. It has been shown that
the presence of the cylinder provides a possibility for the
essential enhancement of the radiation intensity.

In the present paper, on the basis of the Green function obtained
in Ref. \cite{Grigoryan1995}, the radiation from a charged
particle moving along a helical orbit inside a dielectric cylinder
is studied. The corresponding problem for the charge moving in
vacuum has been widely discussed in literature (see, e.g., Refs.
\cite{Soko86,Bord99,Hofm04} and references given therein). This
type of electron motion is used in helical undulators for
generating electromagnetic radiation in a narrow spectral interval
at frequencies ranging from radio or millimeter waves to X-rays
\cite{Alfe74,Kinc77} (see also
\cite{Bord99,Hofm04,Luch90,Nikitin}). The paper is organized as
follows. In next section the vector potential and electromagnetic
fields are determined for the region outside the cylinder by using
Green function. A formula for the angular-frequency distribution
for the radiation intensity is derived and various limiting cases
are investigated in Sec. \ref{sec:radiation}. In Sec.
\ref{sec:Properties} we discuss the characteristic features  of
the radiation intensity and the results of the numerical
evaluations are presented. Furthermore, we demonstrate the
possibility for the appearance of strong narrow peaks in the
angular distribution of the radiation intensity at a given
harmonic. The case of helical undulator with non-relativistic
transversal motion is considered in Sec. \ref{sec:helond}. Various
regimes for the undulator parameter are investigated. Section
\ref{sec:Conc} concludes the main results of the paper.

\bigskip

\section{Electromagnetic potentials and fields in the exterior region}

\label{sec:oscfields}

Consider a point charge $q$ moving along the helical trajectory of
radius $\rho _{0}$ inside a dielectric cylinder with radius $\rho
_{1}$ and with permittivity $\varepsilon _{0}$. We will assume
that this system is immersed in a homogeneous medium with
dielectric permittivity $\varepsilon _{1}$ (for simplicity the
magnetic permeability is taken to be unit). The particle
velocities along the axis of the cylinder (drift velocity) and in
the perpendicular plane we will denote by $v_{\parallel }$ and
$v_{\perp }$, respectively. In a properly chosen cylindrical
coordinate system ($\rho ,\phi ,z$) the corresponding motion is
described by the coordinates
\begin{equation}
\rho =\rho _0,\quad \phi =\omega _0 t,\quad z=v_{\parallel }t,
\label{hetagic}
\end{equation}%
where the $z$-axis coincides with the cylinder axis and $\omega
_{0}=v_{\perp }/\rho _{0}$ is the angular velocity of the charge.
This type of motion can be produced by a uniform constant magnetic
field directed along the axis of a cylinder, by a circularly
polarized plane wave, or by a spatially periodic transverse
magnetic field of constant absolute value and a direction that
rotates as a function of the coordinate $z$.

The solution to Maxwell's equations for the four-vector potential
is expressed in terms of the Green function $G_{il}({\bf r},t,{\bf
r}^{\prime },t^{\prime })$ of the electromagnetic field being a
second-rank tensor:
\begin{equation}
A_{i}({\bf r},t)=-\frac{1}{2\pi ^{2}c}\int G_{il}({\bf r},t,{\bf
r}^{\prime },t^{\prime })j_{l}({\bf r}^{\prime },t)d{\bf
r}^{\prime }dt^{\prime }, \label{vecpot}
\end{equation}%
where the summation over $l$ is understood and the components of
the current density created by the charge are given by formula
\begin{equation}
j_{l}=\frac{q}{\rho }v_{l}\delta (\rho -\rho _{0})\delta (\phi
-\omega _{0}t)\delta (z-v_{\parallel }t). \label{hosqxtutjun}
\end{equation}
For static and cylindrically symmetric medium the Green function
can be developed in a Fourier expansion:
\begin{equation}
G_{il}({\bf r},t,{\bf r}^{\prime }t^{\prime })=\sum_{m=-\infty }^{\infty
}\int_{-\infty }^{\infty }dk_{z}d\omega G_{il}(m,k_{z},\omega ,\rho ,\rho
^{\prime })\exp [im(\phi -\phi ^{\prime })+ik_{z}(z-z^{\prime })-i\omega
(t-t^{\prime })].  \label{GF_furieexp}
\end{equation}%
The substitution of expressions (\ref{hosqxtutjun}) and
(\ref{GF_furieexp}) into formula (\ref{vecpot}) gives the
following result for the components of the vector potential:
\begin{equation}\label{vecpot1}
\begin{split}
A_{l}({\bf r},t) =&-\frac{q}{\pi c}\sum_{m=-\infty }^{\infty }e^
{im(\phi -\omega _{0}t)}\int_{-\infty }^{\infty }dk_{z}e^{
ik_{z}(z-v_{\parallel }t)} \\
& \times \left[ v_{\perp }G_{l\phi }(m,k_{z},\omega
_{m}(k_{z}),\rho ,\rho _{0}) +v_{\parallel } G_{lz}(m,k_{z},\omega
_{m}(k_{z}),\rho ,\rho _{0})\right] ,
\end{split}
\end{equation}%
where
\begin{equation}
\omega _{m}(k_{z})=m\omega _{0}+k_{z}v_{\parallel }.  \label{omegam}
\end{equation}%
Using the expression of the Green function given earlier in Ref.
\cite{Grigoryan1995} and introducing the notations
\begin{equation}
\lambda ^2_{j}=\frac{\omega ^{2}_{m}(k_z)}{c^{2}}\varepsilon
_{j}-k_{z}^{2} ,\quad j=0,1,  \label{lambdaj}
\end{equation}%
for the corresponding Fourier components
$G_{il}=G_{il}(m,k_{z},\omega _{m}(k_{z}),\rho ,\rho _{0})$ in the
region outside the cylinder, $\rho >\rho _{1}$, we obtain
\begin{subequations}\label{GFcomp}
\begin{eqnarray}
G_{\rho \phi } &=&\frac{i}{2}
\sum_{p=\pm 1}pB_{m}^{(p)}H_{m+p}(\lambda _{1}\rho ), \label{GFcompa}\\
G_{\rho z} &=&\frac{ik_{z}}{%
2\rho _{1}}\frac{J_{m}(\lambda _{0}\rho _{0})H_{m}(\lambda
_{1}\rho _{1})}{\alpha _m W_{m}}\sum_{p=\pm
1}\frac{J_{m+p}(\lambda
_{0}\rho _{1})}{W_{m+p}}H_{m+p}(\lambda _{1}\rho ), \label{GFcompb}\\
G_{\phi \phi } &=&\frac{1}{2}
\sum_{p=\pm 1}B_{m}^{(p)}H_{m+p}(\lambda _{1}\rho ), \label{GFcompc}\\
G_{\phi z} &=&\frac{k_{z}}{2\rho _{1}}\frac{J_{m}(\lambda _{0}\rho
_{0})H_{m}(\lambda _{1}\rho _{1})}{\alpha _{m}W_{m}}\sum_{p=\pm 1}p\frac{%
J_{m+p}(\lambda _{0}\rho _{1})}{W_{m+p}}H_{m+p}(\lambda _{1}\rho ), \label{GFcompd}\\
G_{zz} &=&\frac{J_{m}(\lambda _{0}\rho _{0})}{\rho
_{1}W_{m}}H_{m}(\lambda _{1}\rho ), \label{GFcompe}
\end{eqnarray}
\end{subequations}%
where $J_{m}(x)$ is the Bessel function and
$H_{m}(x)=H_{m}^{(1)}(x)$ is the Hankel function of the first
kind. The coefficients $B_{m}^{(p)}$, $p=\pm 1$, in these formulae
are determined by the expressions
\begin{equation}
B_{m}^{(p)}=\frac{1}{\rho _{1}W_{m+p}}\left[ J_{m+p}(\lambda
_{0}\rho _{0})+\frac{p \lambda _{1}}{2\alpha _{m}}J_{m+p}(\lambda
_{0}\rho _{1})H_{m}(\lambda _{1}\rho _{1})\sum_{l=\pm
1}\frac{J_{m+l}(\lambda _{0}\rho _{0})}{W_{m+l}}\right] ,
\label{Bm+p}
\end{equation}%
where
\begin{equation}
\alpha _{m}=\frac{\varepsilon _{0}}{\varepsilon
_{1}-\varepsilon _{0}}-\frac{\lambda _{0}J_{m}(\lambda _{0}\rho _{1})}{2}%
\sum_{l=\pm 1}l\frac{H_{m+l}(\lambda _{1}\rho _{1})}{W_{m+l}}.  \label{bet1}
\end{equation}%
Here and below we use the notation%
\begin{equation}
W_{m}=J_{m}(\lambda _{0}\rho _{1})\frac{\partial H_{m}(\lambda _{1}\rho _{1})%
}{\partial \rho _{1}}-H_{m}(\lambda _{1}\rho _{1})\frac{\partial
J_{m}(\lambda _{0}\rho _{1})}{\partial \rho _{1}}.  \label{Wronskian}
\end{equation}%
In the definition of $\lambda _{1}$ from Eq. (\ref{lambdaj}) one should take
into account that in the presence of the imaginary part $\varepsilon
_{1}^{\prime \prime }(\omega )$ for dielectric permittivity ($%
\varepsilon _{1}=\varepsilon _{1}^{\prime }+i\varepsilon _{1}^{\prime \prime
}$) the radiation field in the exterior medium must damp exponentially for
large $\rho $. This leads to the following relations:
\begin{equation}
\lambda _{1}=\begin{cases} (\omega _m/c)\sqrt{\varepsilon
_{1}-k_{z}^{2}c^{2}/\omega ^{2}_m}, &  \omega
^{2}_m\varepsilon _{1}/c^{2}>k_{z}^{2}, \\
i\sqrt{k_{z}^{2}-\omega ^{2}_m\varepsilon _{1}/c^{2}}, & \omega
^{2}_m\varepsilon _{1}/c^{2}<k_{z}^{2}.
\end{cases} \label{lambda1}
\end{equation}%
Note that for $\lambda _{1}^{2}>0$ the sign of $\lambda _{1}$ may be also
determined from the principle of radiation (different signs of $\omega t$
and $\lambda _{1}\rho $ in the expressions for the fields) for large $\rho $.

Substituting the expressions for the components of the Green
function into formula (\ref{vecpot1}), we present the vector
potential in the form of Fourier expansion
\begin{equation}
A_{l}({\bf r},t)=\sum_{m=-\infty }^{\infty }e^{im(\phi -\omega
_{0}t)}\int_{-\infty }^{\infty }dk_{z}e^{ik_{z}(z-v_{\parallel
}t)}A_{ml}(m,k_{z},\rho ), \label{vecpot2}
\end{equation}%
where the Fourier components $A_{ml}=A_{ml}(m,k_{z},\rho )$ are
determined by formulae
\begin{subequations}\label{vecpot3}
\begin{eqnarray}
A_{m\rho } &=&-\frac{q i}{2\pi }\sum_{p=\pm
1}pC_{m}^{(p)}H_{m+p}(\lambda
_{1}\rho )   \label{vecpot3a}\\
A_{m\phi } &=&-\frac{q i}{2\pi }\sum_{p=\pm
1}C_{m}^{(p)}H_{m+p}(\lambda
_{1}\rho ) \label{vecpot3b}\\
A_{mz} &=&-\frac{q}{\pi c}\frac{v_{\parallel }}{\rho
_{1}W_{m}}J_{m}(\lambda _{0}\rho _{0})H_{m}(\lambda _{1}\rho ),
\label{vecpot3c}
\end{eqnarray}
\end{subequations}%
with the coefficients
\begin{equation}
C_{m}^{(p)}=\frac{v_{\perp }}{c}B_{m}^{(p)}+pv_{\parallel
}k_{z}\frac{J_{m}(\lambda _{0}\rho _{0})J_{m+p}(\lambda _{0}\rho
_{1})H_{m}(\lambda _{1}\rho _{1})}{c \rho {_{1}}\alpha
_{m}W_{m}W_{m+p}}. \label{Cm+p}
\end{equation}%
The strengths of electromagnetic field are obtained by means of
standard formulae of electrodynamics. As is seen from formula
(\ref{vecpot2}), analogous expressions may also be written for the
electric and magnetic fields. Having the vector potential, one can
derive the corresponding Fourier coefficients for the magnetic
field:
\begin{subequations}\label{magnetic}
\begin{eqnarray}
H_{m\rho } &=&\frac{iqk_{z}}{2\pi }\sum_{p=\pm
1}D_{m}^{(p)}H_{m+p}(\lambda
_{1}\rho ),  \label{magnetica}\\
H_{m\phi } &=&\frac{qk_{z}}{2\pi }\sum_{p=\pm
1}pD_{m}^{(p)}H_{m+p}(\lambda
_{1}\rho ),  \label{magneticb}\\
H_{mz} &=&-\frac{q\lambda _{1}}{2\pi }\sum_{p=\pm
1}pD_{m}^{(p)}H_{m}(\lambda _{1}\rho ), \label{magneticc}
\end{eqnarray}
\end{subequations}%
where the notation%
\begin{equation}
D_{m}^{(p)}=C_{m}^{(p)}-\frac{v_{\parallel }\lambda _{1}}{c
k_{z}\rho _{1}}\frac{J_{m}(\lambda _{0}\rho _{0})}{W_{m}},\quad
p=\pm 1,  \label{Dm}
\end{equation}%
is introduced. By making use of Maxwell's equation $\nabla \times
{\bf H}=-i\omega \varepsilon _{1}{\bf E}/c$, one can derive the
corresponding Fourier coefficients for the electric field:
\begin{subequations}\label{electric}
\begin{eqnarray}
E_{m\rho } &=&\frac{qc}{4\pi \omega _{m}(k_z)\varepsilon
_{1}}\sum_{p=\pm 1}pH_{m+p}(\lambda _{1}\rho )\left\{ \left(
\frac{\omega _{m}^{2}\varepsilon _{1}}{c^{2}}+k_{z}^{2}\right)
D_{m}^{(p)}-\lambda _{1}^{2}D_{m}^{(-p)}\right\} ,
 \label{electrica} \\
E_{m\phi } &=&-\frac{q ic}{4\pi \omega _{m}(k_z)\varepsilon
_{1}}\sum_{p=\pm 1}H_{m+p}(\lambda _{1}\rho )\left\{ \left(
\frac{\omega _{m}^{2}\varepsilon _{1}}{c^{2}}+k_{z}^{2}\right)
D_{m}^{(p)}-\lambda _{1}^{2}D_{m}^{(-p)}\right\} ,
\label{electricb} \\
E_{mz} &=&\frac{qic\lambda _{1}k_{z}}{2\pi \omega
_{m}(k_z)\varepsilon _{1}} \sum_{p=\pm 1}D_{m}^{(p)}H_{m}(\lambda
_{1}\rho ). \label{electricc}
\end{eqnarray}
\end{subequations}%
As follows from these formulae, ${\bf E}_{m}\cdot {\bf H}_{m}=0$,
i.e., the Fourier components of the electric and magnetic fields
are perpendicular. In the case of a particle motion in a
homogeneous medium one has $\varepsilon _0=\varepsilon _1$ and the
terms in the definitions of the coefficients $C_m^{(p)}$ and
$B_m^{(p)}$ involving the function $\alpha _m$ vanish. By taking
into account that in this case $W_m=2i/\pi \rho _1$, one finds
\begin{equation}\label{Dmpham}
D_m^{(p)}=\frac{\pi }{2i}\left[ \frac{v_{\perp
}}{c}J_{m+p}(\lambda _1 \rho _0)-\frac{v_{\parallel }\lambda
_1}{ck_z}J_{m}(\lambda _1\rho _0)\right] ,\quad \varepsilon _0=
\varepsilon _1.
\end{equation}

As functions of $k_z$ the Fourier coefficients for the fields
determined by relations (\ref{magnetic}), (\ref{electric}) have
poles corresponding to the zeros of the function $\alpha _m$
appearing in the denominators of Eqs. (\ref{Bm+p}),(\ref{Cm+p}).
It can be seen that this function has zeros only for $\lambda
_1^2<0<\lambda _0^2$. As a necessary condition for this one has
$\varepsilon _1<\varepsilon _0$. Note that for the corresponding
modes the Fourier coefficients are proportional to the MacDonald
function $K_\nu (|\lambda _1|\rho )$, $\nu =m,m\pm 1$, and they
are exponentially damped with the distance from the cylinder axis.
These modes are precisely the eigenmodes of the dielectric
cylinder and propagate inside the cylinder. Below, in the
consideration of the intensity for the radiation to the exterior
medium, we will neglect the contribution of the poles
corresponding to these modes.

\section{Spectral-angular distribution of the radiation
intensity} \label{sec:radiation}

Now we proceed to the consideration of the intensity for the
radiation to the exterior medium. As it follows from the
expressions of the fields, for $\lambda _1^2<0$ the
corresponding Fourier components are exponentially damped for large values $%
\rho $, and the radiation is present only under the condition $\lambda
_{1}^{2}>0$. The average energy flux per unit time through the cylindrical
surface of radius $\rho $ coaxial with the dielectric cylinder is given by
the Poynting vector ${\bf S}$:
\begin{equation}
I=\frac{2\pi }{T}\int_{0}^{T}dt\int_{-\infty }^{\infty }{\bf
S}\cdot {\bf n}_{\rho }\rho dz,\quad {\bf S}=\frac{c}{4\pi }{\bf
E} \times {\bf H},\quad T=\frac{2\pi }{\omega _{0}}.
\label{Poyntingvector}
\end{equation}%
Substituting the corresponding Fourier expansions and using the
fact that the replacement $m\rightarrow -m$, $k_{z}\rightarrow
-k_{z}$ leads to $\omega _{m}(k_{z})\rightarrow -\omega
_{m}(k_{z})$, $\lambda _{1}\rightarrow -\lambda _{1}$,
$H_{m}(-k_{z})=H_{m}^{\ast }(k_{z})$, we obtain
\begin{equation}
I=2c\pi \rho \, {\rm Re}\sum_{m=0}^{\infty }{}^{\prime
}\int_{\lambda _{1}^{2}>0}dk_{z}[E_{m\phi }H_{mz}^{\ast
}-E_{mz}H_{m\phi }^{\ast }], \label{Int1}
\end{equation}%
where the prime over the sum means that the term with $m=0$ must
be taken with the weight $1/2$.  At large distances from the
cylinder, by using the asymptotic expressions of the Hankel
functions, for the radiation intensity one finds
\begin{equation}
I=\frac{q^{2}c^2}{\pi ^{2}}\sum_{m=0}^{\infty }{}^{\prime
}\int_{\lambda _{1}^{2}>0}\frac{dk_{z}}{\varepsilon _{1}|\omega
_{m}(k_z)|}\left[ \frac{\omega _{m}^{2}(k_z)}{c^{2}}\varepsilon
_{1}\left\vert D_{m}^{(+1)}-D_{m}^{(-1)}\right\vert
^{2}+k_{z}^{2}\left\vert D_{m}^{(+1)}+D_{m}^{(-1)}\right\vert
^{2}\right] , \label{Int1ynd}
\end{equation}%
where the coefficients $D_{m}^{(p)}$ are defined by formulae
(\ref{Dm}).

The case $\omega _0=0$ for a fixed $\rho _0$ corresponds to a
charge moving with constant velocity $v_{\parallel }$ on a
straight line $\rho =\rho _0$ parallel to the cylinder axis. For
this case $\omega _m(k_z)=k_zv_{\parallel }$ and Eq.
(\ref{lambdaj}) takes the form $\lambda _{j}^2=k_z^2(\beta
_{j\parallel }^{2}-1)$. Here and below we use the notations
\begin{equation}\label{lambdaj0}
\beta _{j\parallel }=\frac{v_{\parallel }}{c}\sqrt{\varepsilon
_{j}} , \quad \beta _{j\perp }=\frac{v_{\perp
}}{c}\sqrt{\varepsilon _{j}} .
\end{equation}
From the condition $\lambda _{1}^{2}>0$ it follows that the
radiation is present only when the Cherenkov condition is
satisfied for the parallel component of the particle velocity, $
v_{\parallel }$, and dielectric permittivity $\varepsilon _{1}$
for the surrounding medium, where the radiation propagates: $\beta
_{1\parallel }>1$. Introducing the angle $\vartheta $ of the wave
vector with the cylinder axis, from the relation $k_z=\omega
/v_{\parallel }$ we obtain $\cos \vartheta =\beta _{1\parallel
}^{-1}$, i.e., the radiation propagates under the Cherenkov angle
of the external medium. As for the case under consideration
$v_{\perp }=0$, the first term on the right of formula
(\ref{Cm+p}) vanishes and the dependence on $\rho _0$ in the
coefficients $D_{m}^{(p)}$ is in the form of the Bessel function
$J_{m}(\lambda _0 \rho _0)$. It follows from here that in the
limit $\rho _0\to 0$ (particle moves along the axis of the
cylinder) the term with $m=0$ contributes only and from Eq.
(\ref{Int1ynd}) one obtains
\begin{equation}\label{Irho00}
I|_{\rho _0=0 }=\frac{4q^2v_{\parallel }}{\pi ^2\rho _1^2} \int
_{\beta _{1\parallel }>1} d\omega \, \frac{1}{\varepsilon _1
\omega |W_{\varepsilon }|^2} ,
\end{equation}
where $\omega =|k_z| v_{\parallel }$,
\begin{equation}\label{Weps}
W_{\varepsilon }=J_0(\lambda ^{(0)}_0 \rho _1) H_1(\lambda
^{(0)}_1 \rho _1)- \frac{\varepsilon _0 \lambda
^{(0)}_1}{\varepsilon _1 \lambda ^{(0)}_0} J_1(\lambda ^{(0)}_0
\rho _1) H_0(\lambda ^{(0)}_1 \rho _1) ,
\end{equation}
and
\begin{equation}\label{lambjnew}
\lambda ^{(0)}_j=\frac{\omega }{v_{\parallel }}\sqrt{\beta
_{j\parallel }^2-1}.
\end{equation}
It can be checked that expression (\ref{Irho00}) coincides with
the formula presented, for example in Ref. \cite{Bolotovsky}, for
the radiation of a charge moving along the axis of a cylindrical
channel in a dielectric.

Now we turn to the analysis of formula (\ref{Int1ynd}) for the
general case $\omega _0 \neq 0$. First of all we consider the
contribution of the term with $m=0$. For this term $\omega
_{m}(k_z)=k_{z}v_{\parallel }$ and, as in the previous case, from
the condition $\lambda _{1}^{2}>0$ it follows that the radiation
at $m=0$ is present only when $\beta _{1\parallel }>1$. This
radiation propagates under the Cherenkov angle of the external
medium. By using the expressions for the coefficients
$D_{m}^{(p)}$ and introducing a new integration variable $\omega
=|k_{z}|v_{\parallel }$, for the radiation intensity at $m=0$ one
obtains
\begin{equation}
I_{0}=\frac{4q^{2}v_{\parallel }}{\pi ^{2}\rho
_{1}^{2}}\int_{\beta _{1\parallel }>1}\frac{d\omega }{\omega
\varepsilon _{1}}\left[ \beta
_{1\perp }^{2}\frac{\omega ^{2}J_{1}^{2}(\lambda ^{(0)}_{0}\rho _{0})}{%
v_{\parallel }^{2}|W_{1}|^{2}}+\frac{J_{0}^{2}(\lambda ^{(0)}_{0}\rho _{0})}{%
|W_{\varepsilon }|^{2}}\right] ,  \label{Intm0}
\end{equation}%
where $\lambda ^{(0)}_j$ are defined by Eq. (\ref{lambjnew}) and
in the expression for $W_1$ from Eq. (\ref{Wronskian}) we have to
substitute $\lambda _j=\lambda ^{(0)}_j$. In the limit $\rho
_{0}\rightarrow 0$ the first term in the square brackets vanishes
and from Eq. (\ref{Intm0}) we obtain formula (\ref{Irho00}) for
the radiation intensity from a charge uniformly moving along the
axis of a dielectric cylinder. In the case of the charge motion in
a homogeneous medium one has $\varepsilon _0=\varepsilon _1$ and
formula (\ref{Intm0}) for the radiation intensity at $m=0$ takes
the form
\begin{equation}\label{Intm0ham}
I_0^{(0)}=\frac{q^2}{v_{\parallel }} \int_{\beta _{1\parallel }>1}
d\omega \frac{\omega }{\varepsilon _{1}}\left[ \beta _{1\perp }^2
J_{1}^2\left( \lambda ^{(0)}_1 \rho _0 \right) +(\beta
_{1\parallel }^2-1)J_{0}^2\left( \lambda ^{(0)}_1 \rho _0 \right)
\right] .
\end{equation}

Next we consider the radiation intensity at $m\neq 0$ harmonics.
From the condition $\lambda _{1}^{2}>0$ one obtains the following
quadratic inequality with respect to $k_{z}$:
\begin{equation}
k_{z}^{2}(1-\beta _{1\parallel }^{-2})+2k_{z}\frac{m\omega _{0}}{%
v_{\parallel }}+\frac{m^{2}\omega _{0}^{2}}{v_{\parallel }^{2}}>0.
\label{qarakusihav}
\end{equation}%
Let the Cherenkov condition be not fulfilled initially for the
drift velocity of the charge: $\beta _{1\parallel }<1$. In this
case from inequality (\ref{qarakusihav}) we obtain:
\begin{equation}
k_{z}\in \left( -\frac{m\omega _{0}\sqrt{\varepsilon _{1}}}{c(1+\beta
_{1\parallel })},\frac{m\omega _{0}\sqrt{\varepsilon _{1}}}{c(1-\beta
_{1\parallel })}\right) .  \label{kazettirujt}
\end{equation}%
It is convenient to introduce the new variable $\vartheta $ according to
\begin{equation}
k_{z}=\frac{m\omega _{0}}{c}\frac{\sqrt{\varepsilon _{1}}\cos \vartheta }{%
1-\beta _{1\parallel }\cos \vartheta },  \label{kazet}
\end{equation}%
where from relation (\ref{kazettirujt}) it follows that $\vartheta \in
(0,\pi )$. Then from expression (\ref{omegam}) we have
\begin{equation}
\omega _{m}(k_{z})=\frac{m\omega _{0}}{1-\beta _{1\parallel }\cos \vartheta }%
,\ m=1,2\ldots   \label{omegatet}
\end{equation}%
Note that in accordance with Eqs. (\ref{kazet}) and (\ref{omegatet}) the
quantities $k_{z}$ and $\omega _{m}(k_{z})$ are connected by the relation $%
k_{z}=\omega _{m}(k_{z})\sqrt{\varepsilon _{1}}\cos \vartheta \,/c$.

Now consider the case $\beta _{1\parallel }>1$, when the solution of
inequality (\ref{qarakusihav}) has the following form:
\begin{equation}
k_{z}\in \left( -\infty ,-\frac{m\omega _{0}\sqrt{\varepsilon _{1}}}{c(\beta
_{1\parallel }-1)}\right) \cup \left( -\frac{m\omega _{0}\sqrt{\varepsilon
_{1}}}{c(\beta _{1\parallel }+1)},\infty \right) .  \label{kazettirujt1}
\end{equation}%
Introducing again the variable $\vartheta $ according to expression (\ref%
{kazet}), we see that $\vartheta \in (0,\vartheta _{0})\cup
(\vartheta _{0},\pi )$, where $\vartheta _{0}=\arccos (\beta
_{1\parallel }^{-1})$ is the corresponding Cherenkov angle for the
drift velocity. The relation between the variables $k_{z}$,
$\omega _{m}(k_{z})$, $\vartheta $ now is the same as in the case
$\beta _{1\parallel }<1$. At large distances from the charge
trajectory the dependence of elementary waves on the space time
coordinates has the form $\exp [\omega
_{m}(k_{z})\sqrt{\varepsilon _{1}}(\rho \sin \vartheta +z\cos
\vartheta -ct/\sqrt{\varepsilon _{1}})/c]$, which describes the
wave with the frequency
\begin{equation}
\omega _{m}=\left\vert \omega _{m}(k_{z})\right\vert =\frac{m\omega _{0}}{%
|1-\beta {_{1{\parallel }}}\cos \vartheta |},  \label{omegamtet}
\end{equation}%
propagating at the angle $\vartheta $ to the $z$-axis. Formula (\ref%
{omegamtet}) describes the normal Doppler effect in the cases
$\beta _{1\parallel }<1$ and $\beta _{1\parallel }>1$, $\vartheta
>\vartheta _{0}$,
and anomalous Doppler effect in the case $\beta _{1\parallel }>1$,
$\vartheta <\vartheta _{0}$. By making use of the formulae given
above, the expressions for $\lambda _{j}$ can be written as
\begin{subequations} \label{lambond}
\begin{eqnarray}
&& \lambda _{0}=\frac{m\omega _{0}}{c}\frac{\sqrt{\varepsilon
_{0}-\varepsilon _{1}\cos ^{2}\vartheta }}{1-\beta _{1\parallel
}\cos \vartheta },\label{lambonda}\\
&& \lambda _{1}=\frac{m\omega _{0}}{c}\frac{\sqrt{\varepsilon
_{1}}\sin \vartheta }{1-\beta _{1\parallel }\cos \vartheta }.
\label{lambondb}
\end{eqnarray}
\end{subequations}
Introducing instead of $k_{z}$ a new integration variable
$\vartheta $ in accordance with Eq. (\ref{kazet}), from Eq.
(\ref{Int1ynd}) for the radiation intensity at $m\neq 0$ harmonics
one finds
\begin{equation} \label{Int3a}
I_{m\neq 0}=\sum_{m=1}^{\infty }\int d\Omega
\,\frac{dI_{m}}{d\Omega },
\end{equation}
where $d\Omega =\sin \vartheta d\vartheta d\phi $ is the solid
angle element, and
\begin{equation}
\frac{dI_{m}}{d\Omega }=\frac{q^{2}\omega _{0}^{2}m^2\sqrt{\varepsilon _{1}}}{%
2\pi ^{3}c|1-\beta _{1\parallel }\cos \vartheta |^3}\left\{
\left\vert D_{m}^{(1)}-D_{m}^{(-1)}\right\vert ^{2}+\left\vert
D_{m}^{(1)}+D_{m}^{(-1)}\right\vert ^{2}\cos ^{2}\vartheta
\right\} , \label{Int3}
\end{equation}%
is the average power radiated by the charge at a single harmonic
$m$ into a unit solid angle. In the limit $\rho _0 \to 0$ for a
fixed $\omega _0$ the radiation intensity at the harmonic $m\neq
0$ tends to zero as $\rho _0^{2m}$ and the contribution from $m=0$
remains only. The latter corresponds to the Cherenkov radiation
from a charge in uniform motion along the axis of the cylinder.

In the case of a particle moving in a homogeneous medium with
dielectric permittivity $\varepsilon _1$, in formula (\ref{Int3})
one has $\varepsilon _0=\varepsilon _1$ and using formula
(\ref{Dmpham}) for the coefficients $D_m^{(p)}$, we obtain
\begin{equation}\label{Intham}
\frac{dI_{m}^{(0)}}{d\Omega }=\frac{q^2\omega _0^2m^2}{2\pi
c\sqrt{\varepsilon _1}|1-\beta _{1\parallel }\cos \vartheta |^3}
\left[ \beta _{1\perp }^2 J'_{m}{}^2(\lambda _1\rho _0)+\left(
\frac{\cos \vartheta -\beta _{1\parallel }}{\sin \vartheta
}\right) ^2J_m^2(\lambda _1\rho _0)\right] .
\end{equation}
For $\varepsilon _1=1$ this formula is derived in Ref.
\cite{Soko68} (see also Refs. \cite{Soko86,Tern85,Bord99}). In the
case $v_{\parallel }=0$ and for an arbitrary $\varepsilon _1$,
from (\ref{Intham}) we obtain the formula for the synchrotron
radiation in a homogeneous medium (see Refs.
\cite{Tsytovich,Kitao}). For a charge with a purely transversal
motion ($v_{\parallel }=0$) from Eqs. (\ref{Cm+p}), (\ref{Dm}) we
see that $D_m^{(p)}=(v/c)B_m^{(p)}$, and from Eq. (\ref{Int3}) the
formula derived in Ref. \cite{Kota02} is recovered.

From the analysis presented above it follows that in presence of a
dielectric cylinder the total intensity of the radiation can be
presented as the sum
\begin{equation} \label{totintOm}
I=I_{0}+I_{m\neq 0}\, ,
\end{equation}
where the first term on the right is given by formula
(\ref{Intm0}) and describes the radiation with a continuous
spectrum propagating at the Cherenkov angle of the external
medium, if the condition $\beta _{1\parallel }>1$ is fulfilled
(for $\beta _{1\parallel }<1$ this term is absent). The second
term describes the radiation, which for a given angle $\vartheta $
has a discrete spectrum determined by formula (\ref{omegamtet}).
With allowance for the dispersion of the dielectric permittivity,
this term does not contribute to the radiation at the Cherenkov
angle. This is connected with the fact that for a given $m>0$ the
frequency defined by Eq. (\ref{omegamtet}), tends to infinity as
$\vartheta $ approaches $\vartheta _0$ and hence beginning from a
certain frequency the Cherenkov condition ceases to be fulfilled.
The angles for which the dispersion should be taken into account,
are determined implicitly from the condition $\omega _m\geq \omega
_d$ by using formula (\ref{omegamtet}) and frequency dependence of
the permittivity $\varepsilon _1=\varepsilon _1(\omega _m)$, where
$\omega _d$ is the characteristic frequency of the dispersion.

\section{Properties of the radiation} \label{sec:Properties}

In this section, on the base of formula from previous section we
consider characteristic features of the radiation intensity. For a
non-relativistic particle, $\beta _{1\perp }, \beta _{1\parallel
}\ll 1$, from the general formula one obtains
\begin{equation}\label{Intnonrel}
\frac{dI_{m}}{d\Omega }\approx \frac{8q^2 c\sqrt{\varepsilon
_1}}{\pi \rho _0^2(m!)^2 (\varepsilon _1+\varepsilon _0)^2} \left(
\frac{m\beta _{1\perp }}{2}\right) ^{2(m+1)}\left( 1+\cos ^2
\vartheta \right) \sin ^{2(m-1)}\vartheta .
\end{equation}
In this limit the energy in the higher harmonics is small compared
to that in the fundamental, $m=1$.

For the case $\varepsilon _0<\varepsilon _1$ the quantity $\lambda
_0$ is purely imaginary in the angular region $\cos ^2\vartheta
>\varepsilon _0/\varepsilon _1$. By using the asymptotic formulae
for the Bessel modified function for large values of the argument,
it can be easily seen that in this region the radiation intensity
exponentially decreases with increasing $\rho _1$ in the limit
when the wavelength for the radiation is much less than the
cylinder radius:
\begin{equation}\label{Intlargerad}
\frac{dI_m}{d\Omega }\propto e^{-2|\lambda _0|\rho _1}
I_{m}^2(|\lambda _0|\rho _0), \quad |\lambda _0|\rho _1 \gg 1.
\end{equation}
This is caused by the fact that for $\varepsilon _0/\varepsilon
_1<1$ the angle $\arccos (\sqrt{\varepsilon _{0/}\varepsilon _1})$
corresponds to the angle of total internal reflection, and in the
limit of the geometric optics the beams incident from inside on
the cylinder surface cannot propagate at the angles $\cos
^2\vartheta
>\varepsilon _0/\varepsilon _1$ in the external medium.

Now let us consider the behavior of the radiation intensity near
the Cherenkov angle when $|1-\beta _{1\parallel }\cos \vartheta
|\ll 1$. This behavior is radically different for the cases $\beta
_{0\parallel }>1$ and $\beta _{0\parallel }<1$. In the first case
the quantity $\lambda _0$ is real and using the asymptotic
formulae for the cylinder functions for large values of the
argument it can be seen that (assuming $\varepsilon _0\neq
\varepsilon _1$)
\begin{equation}\label{dImasnearCh1}
\frac{dI_m}{d\Omega }\propto |1-\beta _{1\parallel }\cos \vartheta
| ^{-4}, \quad \beta _{0\parallel }>1.
\end{equation}
In the second case, $\beta _{0\parallel }<1$, the quantity
$\lambda _0$ is purely imaginary and the radiation intensity is
exponentially suppressed:
\begin{equation}\label{dImasnearCh2}
\frac{dI_m}{d\Omega }\propto |1-\beta _{1\parallel }\cos \vartheta
| ^{-4}\exp \left[ -\frac{2m\omega _0(\rho _1-\rho
_0)\sqrt{1-\beta _{0\parallel }^2}}{v_{\parallel }|1-\beta
_{1\parallel }\cos \vartheta |}\right], \quad \beta _{0\parallel
}<1.
\end{equation}
In the same limit, $|1-\beta _{1\parallel }\cos \vartheta |\ll 1$,
for the intensity of radiation in a homogeneous medium from
formula (\ref{Intham}) one has
\begin{equation}\label{dImasnearChham}
\frac{dI_m^{(0)}}{d\Omega }\propto |1-\beta _{1\parallel }\cos
\vartheta | ^{-2}.
\end{equation}
Comparing with Eq. (\ref{dImasnearCh1}), we see that in this limit
the presence of the dielectric cylinder essentially increases the
radiation intensity. Note that in the considered limit, in
accordance with Eq. (\ref{omegamtet}), the frequency for the
radiation is large and the dispersion of the dielectric
permittivity may be important.

\begin{figure}[tbph]
\begin{center}
\begin{tabular}{cc}
\epsfig{figure=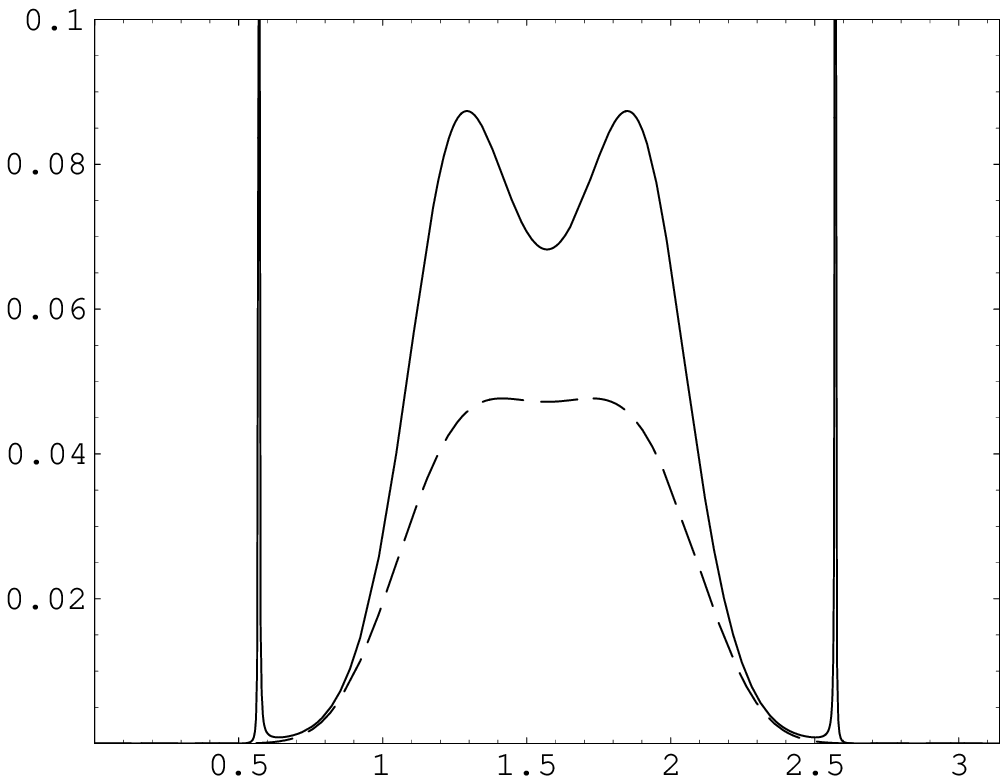,width=6.5cm,height=5.5cm} & \quad %
\epsfig{figure=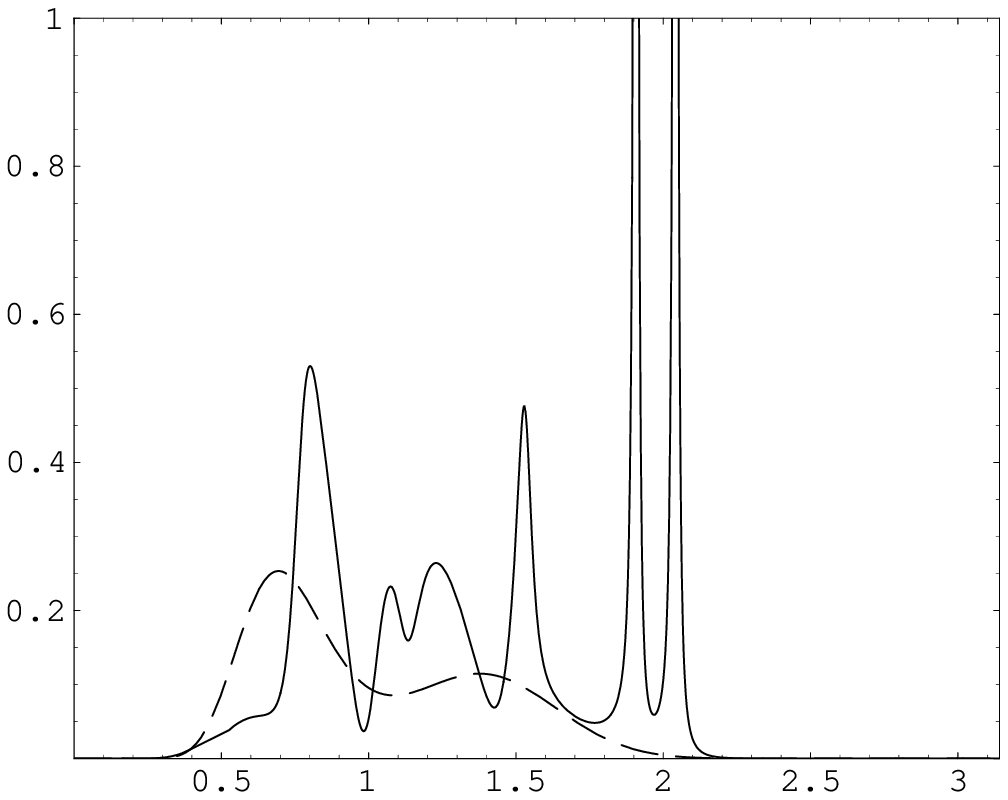,width=6.5cm,height=5.5cm}%
\end{tabular}%
\end{center}
\caption{The dependence of the angular density for the number of
radiated quanta, $(T\sqrt{\varepsilon _1}\hbar c/q^2)dN_m/d\Omega
$, per period $T$ of charge revolution as a function on the
opening angle $\vartheta $ for $\beta _{1\perp }=0.9$, $\rho
_1/\rho _0=1.05$, $m=10$. The left panel is plotted for $\beta
_{1\parallel }=0$ and the right panel is plotted for $\beta
_{1\parallel }=0.5$. Full and dashed curves correspond to the
cases $\varepsilon _0/\varepsilon _1=3$ and  $\varepsilon
_0/\varepsilon _1=1$ respectively. } \label{fig1}
\end{figure}

In Figure \ref{fig1} we have plotted the dependence of the angular
density for the number of the radiated quanta
\begin{equation}\label{dNm}
\frac{dN_m}{d\Omega }=\frac{1}{\hbar \omega _{m}}
\frac{dI_m}{d\Omega },
\end{equation}
as a function on the angle $0\leq \vartheta \leq \pi $ for $\beta
_{1\perp }=0.9$, $\rho _1/\rho _0 =1.05$, $m=10$. The full and
dashed curves correspond to the cases $\varepsilon _0/\varepsilon
_1=3$ and $\varepsilon _0/\varepsilon _1=1$ (homogeneous medium)
respectively. The left panel is plotted for $\beta _{1\parallel
}=0$ (purely transversal motion, the curves are symmetric with
respect to the rotation plane $\vartheta =\pi /2$) and the right
panel is for $\beta _{1\parallel }=0.5$. In both graphics the
strong narrow peaks appear for the case when the dielectric
cylinder is present. At the peaks of the left panel one has
$(T\sqrt{\varepsilon _1}\hbar c/q^2)dN_m/d\Omega \approx 1.6\cdot
10^4$ and the width of the peaks is of the order $\Delta \vartheta
\approx 6\cdot 10^{-5}$. For the left peak on the right panel one
has the value $\approx 4.8$ at the peak and $\Delta \vartheta
\approx 0.01$ and for the right peak one has the value $\approx
11$ at the peak with the width $\Delta \vartheta \approx 0.008$.

\begin{figure}[tbph]
\begin{center}
\begin{tabular}{cc}
\epsfig{figure=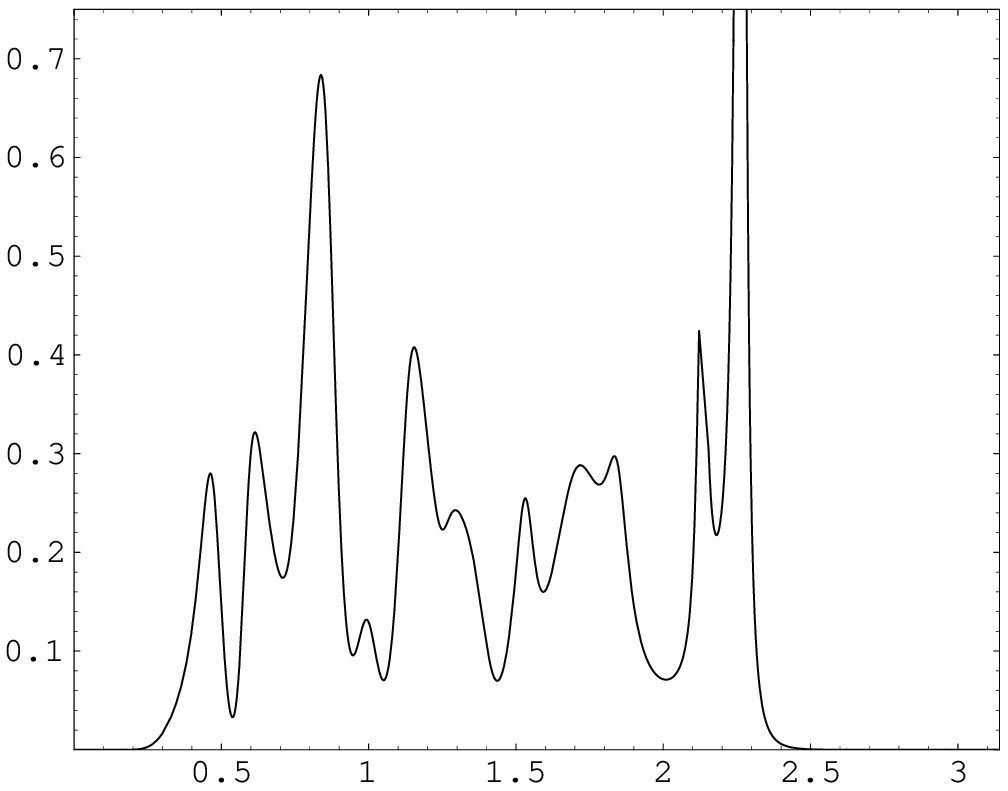,width=6.5cm,height=5.5cm} & \quad %
\epsfig{figure=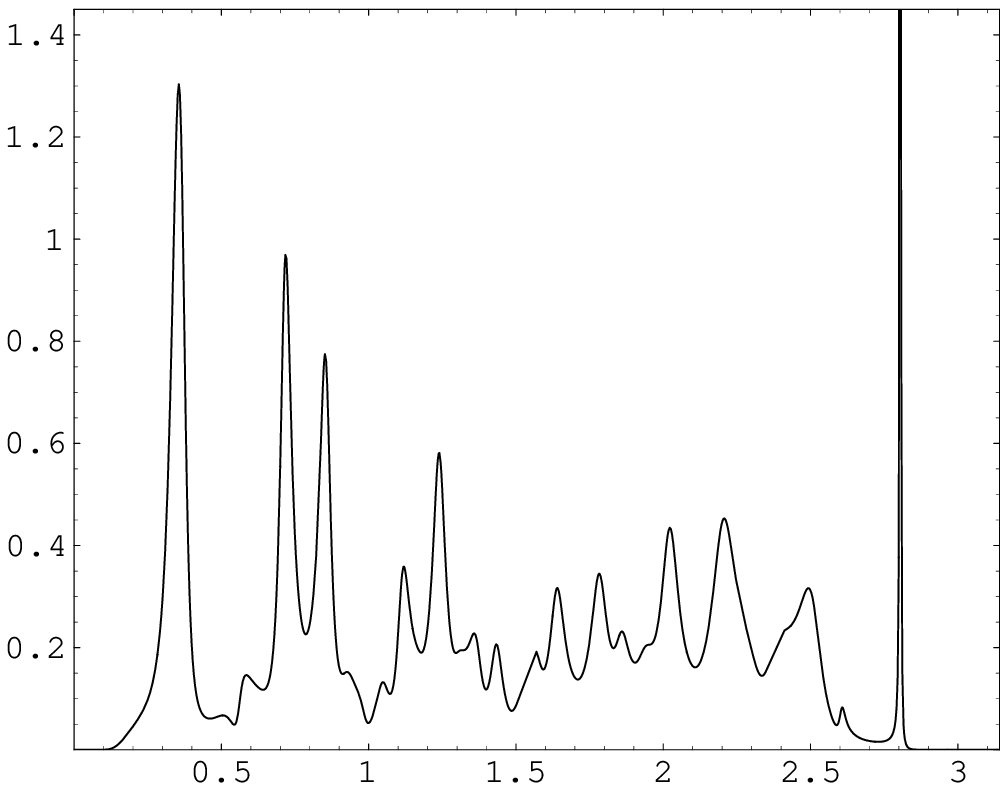,width=6.5cm,height=5.5cm}%
\end{tabular}%
\end{center}
\caption{The dependence of the quantity $(T\sqrt{\varepsilon
_1}\hbar c/q^2)dN_m/d\Omega $ on the angle $\vartheta $ for $\rho
_1/\rho _0=1.5$ (left panel) and $\rho _1/\rho _0=3$ (right
panel). The values of the other parameters are $\beta _{1\perp
}=0.9$, $m=10$, $\beta _{1\parallel }=0.5$, $\varepsilon
_0/\varepsilon _1=3$. } \label{fig2}
\end{figure}

Figure \ref{fig2} presents the angular dependence of the number of
the radiated quanta for the values $\rho _1/\rho _0=1.5$ (left
panel) and $\rho _1/\rho _0=3$ (right panel). The values of the
other parameters are $\beta _{1\perp }=0.9$, $m=10$, $\beta
_{1\parallel }=0.5$, $\varepsilon _0/\varepsilon _1=3$. For the
left panel the value of the function at the peak corresponding to
$\vartheta \approx 2.27$ is equal $\approx 2.5$. For the left
panel the peak corresponds to $\vartheta \approx 2.81$ with the
value of the function $\approx 11.5$.

To understand the presence of the strong narrow peaks in the
angular distribution of the radiation intensity note that for
large $m$ from Debye's asymptotic expansion one has the following
asymptotic formulae for the Bessel and Neumann functions
\begin{equation}\label{JmYmas}
\begin{split}
J_{m}(my)&\sim ({\mathrm{sign}\, y})^{m}\frac{\exp \left[ -m\zeta
(y)\right] }{\sqrt{2\pi
m}(1-y^2)^{1/4}},\quad |y|<1, \\
Y_{m}(my)&\sim ({\mathrm{sign}\, y})^{m+1}\frac{2\exp \left[
m\zeta (y)\right] }{\sqrt{2\pi m}(1-y^2)^{1/4}},\quad |y|<1,
\end{split}
\end{equation}
with
\begin{equation}\label{zeta1}
\zeta (y)=\ln \frac{1+\sqrt{1-y^2}}{|y|}-\sqrt{1-y^2}.
\end{equation}
For $|y|>1$ the corresponding formulae are obtained replacing the
exponential functions by $\sin $ and $\cos $ (see, for instance,
\cite{Abramovic}). From formulae (\ref{JmYmas}) it follows that
for $|y|<1$ the ratio $J_{m}(my)/Y_{m}(my)$ is exponentially small
for large values $m$. Assuming that $|\lambda _1|\rho _1 <m$ and
expanding over the small ratio $J_{m}(\lambda _1\rho
_1)/Y_{m}(\lambda _1\rho _1)$, for the coefficient $\alpha _{m}$
defined by Eq. (\ref{bet1}), one obtains
\begin{equation}\label{ro1bet1}
\begin{split}
\alpha _m \approx & \frac{\varepsilon _0}{\varepsilon
_1-\varepsilon _0}-\frac{1}{2}\sum _{l=\pm 1}F_m^{(l)}\left\{
1-i\left[  \frac{J_{m+l}(\lambda _0 \rho
_1)}{Y_{m+l}(\lambda _1 \rho _1)}\right. \right. \\
& - \left. \left.\frac{\lambda _1J_{m+l}(\lambda _0 \rho
_1)J_{m}(\lambda _1 \rho _1)-\lambda _0J_{m}(\lambda _0 \rho
_1)J_{m+l}(\lambda _1 \rho _1)}{\lambda _1J_{m+l}(\lambda _0 \rho
_1)Y_{m}(\lambda _1 \rho _1)-\lambda _0J_{m}(\lambda _0 \rho
_1)Y_{m+l}(\lambda _1 \rho _1)}\right] \right\} ,
\end{split}
\end{equation}
where
\begin{equation}\label{Eml}
F_m^{(l)}=\left[ \frac{\lambda _1}{\lambda
_{0}}\frac{J_{m+l}(\lambda _0\rho _1)Y_{m}(\lambda _1\rho
_1)}{J_{m}(\lambda _0\rho _1)Y_{m+l}(\lambda _1\rho _1)}-1\right]
^{-1},
\end{equation}
and the second summand in the figure braces is exponentially small
for large $m$. From here it follows that, at points where the real
part of the function $\alpha _m$ is equal to zero, the
contribution of the imaginary part into the coefficients
$D_{m}^{(p)}$ can be exponentially large. The corresponding
condition for the real part to be zero has the form
\begin{equation}\label{peakscond}
\sum _{l=\pm 1}F_m^{(l)}=\frac{2\varepsilon _0}{\varepsilon
_1-\varepsilon _0}.
\end{equation}
Note that this equation is obtained from the equation determining
the eigenmodes for the dielectric cylinder by the replacement $H_m
\to Y_m$.

For the further analysis it is convenient to rewrite equation
(\ref{peakscond}) in the form
\begin{equation}\label{peakscond2}
\begin{split}
\left( \lambda _0 \frac{Y'_{m}(\lambda _1\rho _1)}{Y_{m}(\lambda
_1\rho _1)} -\lambda _1 \frac{J'_{m}(\lambda _0\rho
_1)}{J_{m}(\lambda _0\rho _1)} \right) & \left( \lambda _0
\frac{\varepsilon _1}{\varepsilon _0} \frac{Y'_{m}(\lambda _1\rho
_1)}{Y_{m}(\lambda _1\rho _1)} -\lambda _1 \frac{J'_{m}(\lambda
_0\rho _1)}{J_{m}(\lambda _0\rho _1)} \right) = \\
& =\frac{m^2}{\rho _1^2}\left( 1-\frac{\lambda _0^2}{\lambda
_1^2}\right) \left( \frac{\lambda _1^2}{\lambda
_0^2}-\frac{\varepsilon _1}{\varepsilon _0} \right) .
\end{split}
\end{equation}
First consider the case $\lambda _0^2<0$ which is possible only
for $\varepsilon _0<\varepsilon _1$. Introducing instead of
function $J_m(\lambda _0\rho _1)$ the Bessel modified function
$I_m(|\lambda _0|\rho _1)$ and making use the corresponding
uniform asymptotic expansions for large values of the order, we
can solve equation (\ref{peakscond2}) with respect to $\varepsilon
_0/\varepsilon _1$ for given $\lambda _0$ and $\lambda _1$. As a
result we obtain that $\varepsilon _1<\varepsilon _0$ which is in
contradiction with the condition $\lambda _0^2<0$. Hence, the
above mentioned possibility for the appearance of peaks is not
realized for the case $\lambda _0^2<0$ which corresponds to the
angular region $\cos ^2 \vartheta >\varepsilon _0/\varepsilon _1$.
For this reason we will concentrate on the case $\cos ^2 \vartheta
<\varepsilon _0/\varepsilon _1$.

Let us consider separate cases.

a) Let firstly $(\lambda _0\rho _0/m)^2>1$ and, hence, $(\lambda
_0\rho _1/m)^2>1$. By taking into account the condition $|\lambda
_1|\rho _1<m$, assumed earlier, we obtain that $\lambda _1^2<
\lambda _0^2$ and, hence, $\varepsilon _0>\varepsilon _1$. For the
solutions to equation (\ref{peakscond}) the contribution from the
first summand in the figure braces of Eq. (\ref{ro1bet1})
vanishes. By taking into account asymptotics (\ref{JmYmas}), for
the angular density of the number of the radiated quanta we obtain
the following estimate
\begin{equation}\label{dNmest1}
\frac{dN_m}{d\Omega }\propto \exp \left[ 2m \zeta (\lambda _1 \rho
_1/m)\right] ,
\end{equation}
with the function $\zeta (y)$ from Eq. (\ref{zeta1}). As this
function is monotonic decreasing the exponent for a given $m$ is
larger for smaller values of $|\lambda _1| $.

b) Let now $(\lambda _0\rho _0/m)^2<1$ and $(\lambda _0\rho
_1/m)^2>1$. The substitution of the asymptotic formulae
(\ref{JmYmas}) into Eq. (\ref{Int3}) shows that, the peaks arise
only under the condition $(\lambda _1\rho _1)^2<(\lambda _0\rho
_0)^2$, which is possible only for $\varepsilon _0>\varepsilon
_1$. For the angular density of the radiated quanta at the peaks
one obtains
\begin{equation}\label{dNmest2}
\frac{dN_m}{d\Omega }\propto \exp \left\{ 2m \left[ \zeta
(\lambda _1 \rho _1/m)-\zeta (\lambda _0 \rho _0/m)\right]\right\}
.
\end{equation}

c) In the case $(\lambda _0\rho _1/m)^2<1$ we have also $(\lambda
_0\rho _0/m)^2<1$. Again, using the asymptotic formulae
(\ref{JmYmas}), it can be seen that in this case Eq.
(\ref{peakscond2}) has no solutions and the peaks are absent.

Therefore, as necessary conditions for the presence of the strong
narrow peaks in the angular distribution for the radiation
intensity one has $|\lambda _1|\rho _1<m<|\lambda _0|\rho _1$, or
in terms of the angle $\vartheta $
\begin{equation}\label{peakscond3}
\frac{\omega _0\rho _1}{c}\sqrt{\varepsilon _1} \sin \vartheta <
|1-\beta _{1\parallel }\cos \vartheta | < \frac{\omega _0\rho
_1}{c}\sqrt{\varepsilon _0 -\varepsilon _1 \cos ^2 \vartheta }.
\end{equation}
From these conditions it follows that we should have
\begin{equation}\label{peakscond4}
\varepsilon _0>\varepsilon _1, \quad \tilde v\sqrt{\varepsilon
_0}/c >1,
\end{equation}
where $\tilde v=\sqrt{v_{\parallel }^2+ \omega _0^2\rho _1^2}$ is
the velocity of the charge image on the cylinder surface. In
particular, the second condition in Eq. (\ref{peakscond4}) tells
that the Cherenkov condition should be satisfied for the velocity
of the charge image on the cylinder surface and dielectric
permittivity of the cylinder. When the Cherenkov condition for the
velocity of the charge image and dielectric permittivity of the
surrounding medium is not satisfied, $\tilde v \sqrt{\varepsilon
_1}/c <1$, the possible strong peaks are located in the angular
regions defined by the inequality
\begin{equation}\label{peakregion1}
\left| \frac{\tilde v}{c}\sqrt{\varepsilon _1}\cos \vartheta
-\frac{v_{\parallel }}{\tilde v}\right| <\frac{\rho _1v_{\perp
}}{\rho _0 \tilde v}\sqrt{\frac{\tilde v^2}{c^2}\varepsilon _0
-1}.
\end{equation}
If Cherenkov condition $\tilde v \sqrt{\varepsilon _1}/c >1$ is
satisfied, in addition to this inequality the condition
\begin{equation}\label{peakregion2}
 \left| \frac{\tilde v}{c}\sqrt{\varepsilon _1}\cos \vartheta
-\frac{v_{\parallel }}{\tilde v}\right| >\frac{\rho _1v_{\perp
}}{\rho _0 \tilde v}\sqrt{\frac{\tilde v^2}{c^2}\varepsilon _1 -1}
\end{equation}
has to be satisfied. By the presented arguments we can also
estimate the width for the peaks. Taking into account
(\ref{ro1bet1}) and expanding $\alpha _m$ near the angle
corresponding to the peak, $\vartheta =\vartheta _p$, it can be
seen that the angular dependence of the radiation intensity near
the peak has the form
\begin{equation}\label{dImnear}
\frac{dI_m}{d\Omega }\sim \frac{1}{(\vartheta -\vartheta
_p)^2/b_p^2+1} \left( \frac{dI_m}{d\Omega }\right) _{\vartheta
=\vartheta _p},
\end{equation}
where
\begin{equation}\label{b1sim}
b_p\sim J_{m+l}(\lambda _1 \rho _1)/Y_{m+l}(\lambda _1 \rho _1)
|_{\vartheta =\vartheta _p}\sim \exp [-2m \zeta (\lambda _1 \rho
_1/m)] .
\end{equation}
It follows from here that the width of the peak has an order
\begin{equation}\label{Deltatheta}
\Delta \vartheta \sim \exp[-2m\zeta (\lambda _1 \rho _1/m)] .
\end{equation}

As we see from estimates (\ref{dNmest1}), (\ref{dNmest2}), at the
peaks the angular density for the number of emitted quanta
exponentially increases with increasing $m$. However, one has to
take into account that in the realistic situations the growth of
the radiation intensity is limited by several factors. In
particular, the factor which limits the increase, is the imaginary
part for the dielectric permittivity $\varepsilon '' _j$, $j=0,1$.
This leads to an additional summand in the denominator of formula
(\ref{dImnear}), proportional to the ratio $\varepsilon ''
_j/\varepsilon ' _j$, where $\varepsilon ' _j$ is the real part of
the dielectric permittivity. As a result for $b_p^2<\varepsilon ''
_j/\varepsilon ' _j$ the intensity and the width of the peak is
determined by this summand and the saturation of the radiation
intensity at the peak takes place.

\section{Helical undulator}\label{sec:helond}

\begin{figure}[tbph]
\begin{center}
\begin{tabular}{cc}
\epsfig{figure=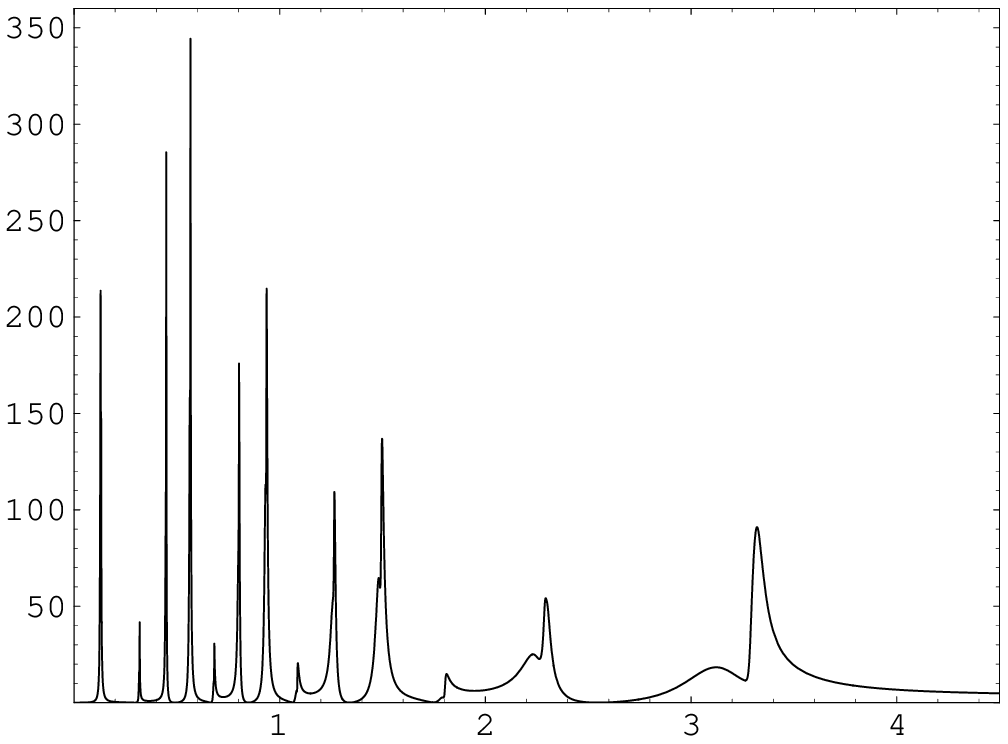,width=6.5cm,height=5.5cm} & \quad %
\epsfig{figure=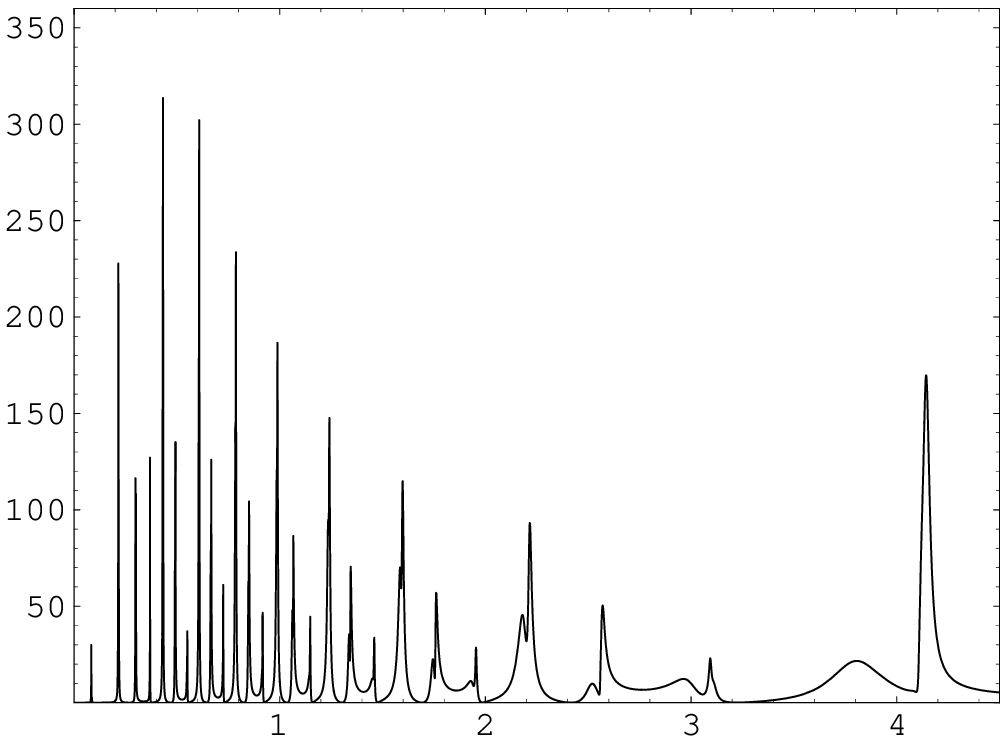,width=6.5cm,height=5.5cm}%
\end{tabular}%
\end{center}
\caption{The dependence of the quantity $(T\sqrt{\varepsilon
_1}\hbar c/q^2)dN_m/d\vartheta d\phi $ on the reduced angle $\chi
$ for $m=1$ (left panel) and $m=2$ (right panel). The values of
the other parameters are $\gamma _{1\parallel }=10$, $\rho _1/\rho
_0=1.5$, $\kappa =1$, $\varepsilon _0/\varepsilon _1=3$. }
\label{fig3}
\end{figure}

Let us consider an important case relativistic charge motion in
direction of the cylinder axis when the velocity of orthogonal
motion is non-relativistic, $\beta _{1\perp }\ll 1$. In helical
undulators this type of motion is realized by a spatially periodic
transverse magnetic field of constant absolute value and a
direction that rotates as a function of the longitudinal
coordinate $z$. In the discussion below we will assume that
$0<1-\beta _{1\parallel }\ll 1$. Under this condition, as it
follows from Eq. (\ref{Int3}), the radiation intensity is sharply
peaked for opening angles near $\vartheta \sim 1/\gamma
_{1\parallel }$, where $\gamma _{1\parallel }=1/\sqrt{1-\beta
_{1\parallel }^2}$ is the longitudinal Lorentz factor. It is
convenient to introduce new variables
\begin{equation}\label{chikappa}
\chi =\gamma _{1\parallel }\vartheta ,\quad \kappa =\beta _{1\perp
}\gamma _{1\parallel },
\end{equation}
in terms of which for small angles $\vartheta $ one has
\begin{equation}\label{1-bet}
1-\beta _{1\parallel }\cos \vartheta \approx \frac{1+\chi ^2}{2
\gamma _{1\parallel }^2}, \quad d\Omega =\frac{\chi }{\gamma
_{1\parallel }^2} d\chi d\phi .
\end{equation}
The parameter $\kappa $ is the so-called undulator parameter which
characterizes the relation between velocity deviation angle and
the opening of the radiation cone. The cases $\kappa
>1$ and $\kappa <1$ correspond to wigglers and undulators,
respectively (strong and weak undulators in another terminology).
For the quantities $\lambda _1\rho _0$ and $\lambda _0\rho _0$ we
obtain
\begin{equation}\label{hellamb10}
\lambda _1\rho _0\approx \frac{2m\kappa \chi }{1+\chi ^2},\quad
\lambda _0\rho _0 \approx \frac{2m\kappa \gamma _{1\parallel
}}{1+\chi ^2} \sqrt{\varepsilon _0/\varepsilon _1-1}.
\end{equation}
On the base of these relations three regimes may be distinguished.

a) The first one corresponds to the case $\kappa \gtrsim 1$ when
$\lambda _1\rho _0 \gtrsim 1$ and $\lambda _0\rho _0\gg 1$. In the
definition of the coefficients $D_{m}^{(p)}$ the Bessel functions
$J_{\nu }(\lambda _0 \rho _j)$ can be replaced by their asymptotic
expansions for large values of the argument (we assume that $\rho
_0 $ and $\rho _1$ are of the same order). The main contribution
into the coefficient $D_{m}^{(p)}$ comes from the second summand
on the right of formula (\ref{Cm+p}). This term dominates the
contribution coming from other terms in Eqs. (\ref{Cm+p}),
(\ref{Dm}) by the factor $\gamma _{1\parallel }$. It can be seen
that the radiation intensity as a function on $\vartheta $ has
sharp peaks at the points $\vartheta =\vartheta _l$ corresponding
to the zeros of the Bessel function $J_{m}(\lambda _0 \rho _1)$,
i.e. at the angles defined from $\lambda _0\rho _1=j_{m,l}$, where
$j_{m,l}$ is the $l$th positive zero of the function $J_{m}(z)$.
Note that as $\lambda _0 \rho _1\gg 1$, for the zeros $j_{m,l}$ we
can use McMahon's expansion for large zeros (see, for instance,
\cite{Abramovic}). For the radiation intensity at the
corresponding peaks one has an estimate $dI_1/d\vartheta d \phi
\propto \gamma _{1\parallel }^3$. At the angles away from the
peaks the radiation intensity is relatively suppressed by the
factor $\gamma _{1\parallel }$: $dI_1/d\vartheta d \phi \propto
\gamma _{1\parallel }^2$. This is illustrated in Figure \ref{fig3}
where the number of radiated quanta at harmonics $m=1$ and $m=2$
is presented as a function on the variable $\chi $ for the value
of the undulator parameter $\kappa =1$. We verified numerically
that the sharp peaks in these graphs are located at the angles
corresponding to the zeros of the function $J_{m}(\lambda _0 \rho
_1)$. In this regime the particle is a relativistic one in the
reference frame moving with the velocity $v_{\parallel }$ and the
spectrum contains many harmonics.

b) In the second regime one has $\kappa \gamma _{1\parallel }\sim
1$ which corresponds $\lambda _1\rho _0\ll 1$, $\lambda _0\rho
_0\sim 1$. In this case we replace the Hankel functions in the
definitions of the coefficients $D_{m}^{(p)}$ by their asymptotic
expressions for small values of the argument. This leads to the
estimate $dI_1/d\vartheta d \phi \propto \gamma _{1\parallel }/\ln
^2 \gamma _{1\parallel }$ and $dI_m/d\vartheta d \phi \propto
1/\gamma _{1\parallel }^{2m-3}$ for the harmonics $m\geq 2$. As we
see in this case the main contribution comes from the fundamental
harmonic $m=1$. This is a simple consequence of the fact that
under the condition $k\ll 1$, the particle has a non-relativistic
velocity in the reference frame which moves with the velocity
$v_{\parallel }$.

c) The third regime corresponds to $\kappa \ll \gamma _{1\parallel
}^{-1}$ when both quantities $\lambda _1 \rho _0$ and $\lambda _0
\rho _0$ are small. Expanding the cylindrical functions, we can
see that, as in the previous case, the main part of the radiation
occurs at the $m=1 $ harmonic and to the leading order one
receives
\begin{equation}\label{dI1ond}
dI_{1}\approx \frac{8q^2 c (\kappa \gamma _{1\parallel })^{4}}{\pi
\rho _0^2 \varepsilon _1^{3/2}} \left( \frac{\varepsilon _0
-\varepsilon _1}{\varepsilon _0 +\varepsilon _1} \right)
^2\frac{\chi }{(1+\chi ^2)^5} d \chi d\phi .
\end{equation}
For the higher harmonics one has an estimate $dI_m/d\vartheta d
\phi \propto (\kappa \gamma _{1\parallel })^{2m+2}/\gamma
_{1\parallel }^{2m-3}$, $m\geq 2$. Integrating over $\chi $ and
$\phi $, we find the total power of radiation in this regime:
\begin{equation}\label{I1ond}
I_1 \approx \frac{2q^2 c}{\rho _0^2 \varepsilon _1^{3/2}} (\kappa
\gamma _{1\parallel })^4 \left( \frac{\varepsilon _0 -\varepsilon
_1}{\varepsilon _0 +\varepsilon _1} \right) ^2 .
\end{equation}
As in the case a), in regimes b) and c) the main contribution into
the radiation intensity  comes from the second term on the right
of formula (\ref{Cm+p}).

For a particle moving in a homogeneous medium, $\varepsilon
_0=\varepsilon _1$, the leading term vanishes and it is necessary
to take the next terms in the corresponding asymptotic expansion.
For small values of the undulator parameter, $\kappa \ll 1$, the
radiation intensity is dominated by the basic harmonic and the
corresponding angular distribution is given by the formula (see,
for example, Ref. \cite{Bord99} for the case $\varepsilon _1=1$)
\begin{equation}\label{dI10ond}
dI_{1}^{(0)}\approx \frac{2q^2 c \kappa ^{4}}{\pi \rho _0^2
\varepsilon _1^{3/2}} \frac{1+\chi ^4}{(1+\chi ^2)^5} \chi d \chi
d\phi .
\end{equation}
Integrating over the angles we obtain the total power of the
radiation in a homogeneous medium:
\begin{equation}\label{I10ond}
I_1^{(0)} \approx \frac{2q^2 c}{3\rho _0^2 \varepsilon _1^{3/2}}
\kappa ^4.
\end{equation}
Now comparing formulae (\ref{I1ond}) and (\ref{I10ond}), we see
that
\begin{equation}\label{I1I10}
\frac{I_1}{I_{1}^{(0)}} \approx 3 \gamma _{1\parallel }^4 \left(
\frac{\varepsilon _0 -\varepsilon _1}{\varepsilon _0 +\varepsilon
_1} \right) ^2,
\end{equation}
and the presence of the cylinder essentially increases the
radiated power to compared with the radiation in the homogeneous
medium.

\begin{figure}[tbph]
\begin{center}
\begin{tabular}{cc}
\epsfig{figure=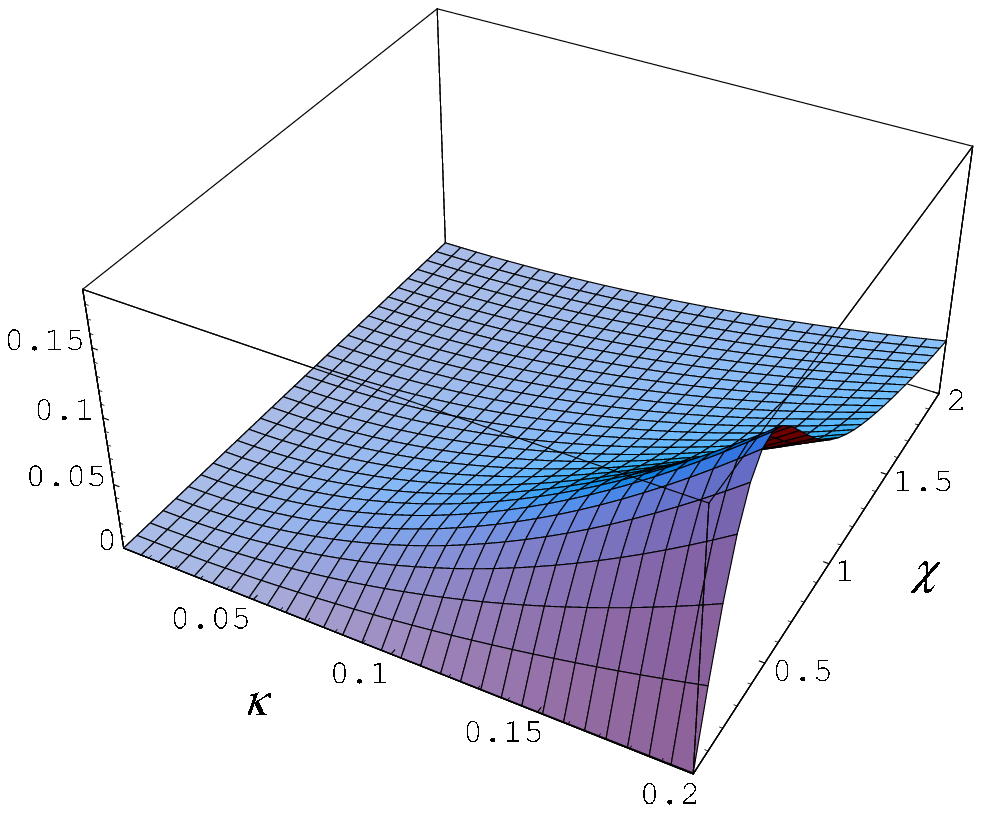,width=6.5cm,height=5.5cm} & \quad %
\epsfig{figure=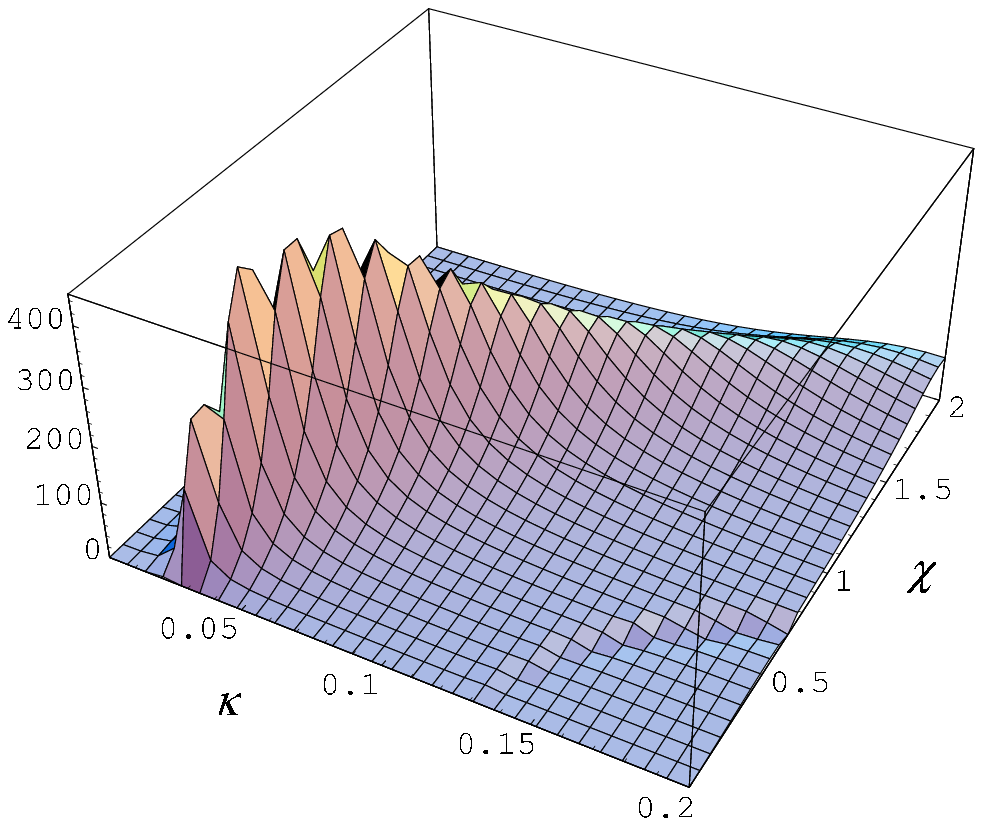,width=6.5cm,height=5.5cm}%
\end{tabular}%
\end{center}
\caption{The dependence of the quantity $(T\sqrt{\varepsilon
_1}\hbar c/q^2)dN_m/d\vartheta d\phi $ on the undulator parameter
$\kappa $ and the reduced angle $\chi $ for $\varepsilon
_0=\varepsilon _1$ (left panel) and $\varepsilon _0=3\varepsilon
_1$ (right panel). The values of the other parameters are $\gamma
_{1\parallel }=10$, $m=1$, $\rho _1/\rho _0=1.05$. } \label{fig4}
\end{figure}

The results of the numerical evaluations of the radiation
intensity for small values of the undulator parameter are
presented in Figure \ref{fig4}, where the function
$(T\sqrt{\varepsilon _1}\hbar c/q^2)dN_m/d\vartheta d\phi $ is
plotted versus the undulator parameter $\kappa $ and the reduced
angle $\chi $ for the values of parameters $\gamma _{1\parallel
}=10$, $m=1$, $\rho _1/\rho _0=1.05$. The left panel corresponds
to the radiation from a charge moving in a homogeneous medium
($\varepsilon _0=\varepsilon _1$) and the right panel corresponds
to the presence of the cylinder with dielectric permittivity
$\varepsilon _0=3\varepsilon _1$.

\section{Conclusion}

\label{sec:Conc}

Synchrotron radiation has become one of the most valuable and
useful scientific tools with ever increasing applications for
basic and applied research. In the present paper we have
considered the influence of a dielectric cylinder on the
synchrotron radiation from a charged particle moving along a
helical orbit inside a dielectric cylinder. By using the Green
function, we have derived formulae for the electromagnetic
potentials and fields, Eqs. (\ref{vecpot3}), (\ref{magnetic}),
(\ref{electric}), and for the angular-frequency distribution of
the radiation intensity emitted into the exterior medium. The
latter is a sum of two terms. The first one, given by formula
(\ref{Intm0}), is present when the Cherenkov condition for
dielectric permittivity of the exterior medium and drift velocity
of the charge is satisfied, $\beta _{1\parallel} >1 $, and
describes the radiation with a continuous spectrum propagating
under the Cherenkov angle of the external medium. The second term
in the total radiation intensity is given by formula
(\ref{Int3a}), where the separate terms (\ref{Int3}) describe the
radiation, which for a given propagation direction has a discrete
spectrum determined by formula (\ref{omegamtet}). We have
investigated various limiting cases of the general formula for the
radiation intensity and have compared this formula to previously
known special cases (radiation in vacuum, purely transversal
synchrotron radiation, radiation from a charge moving with
constant velocity along the axis parallel to the cylinder axis).
In Section \ref{sec:Properties} we have investigated
characteristic features of the radiation. Our numerical
calculations have shown that under certain conditions strong
narrow peaks appear in the angular distribution of the radiation
intensity. By using asymptotic formulae for the cylindrical
functions, we analytically argued the possibility for the presence
of such peaks and the corresponding necessary conditions are
specified. In particular, the peaks are present only when
dielectric permittivity of the cylinder is greater than the
permittivity for the surrounding medium and the Cherenkov
condition is satisfied for the velocity of charge image on the
cylinder surface and the dielectric permittivity of the cylinder.
We have also estimated the width of these peaks. The peaks
correspond to the opening angles for which the real part of the
function $\alpha _m$ from (\ref{bet1}) vanishes. The corresponding
condition is in form (\ref{peakscond}) and is obtained from the
equation determining the eigenmodes for the dielectric cylinder
replacing the Hankel functions by the Neumann ones. In section
\ref{sec:helond} an application of the general formula is
considered to the practically important case of the
non-relativistic transverse motion when the longitudinal motion is
relativistic. This type of motion can be produced by a variety of
electromagnetic field structures and is used in helical undulators
for generating electromagnetic radiation in a narrow spectral
interval. Different regimes for the undulator parameter are
considered corresponding to strong and weak undulators. In the
former case the radiation intensity contains many harmonics and
has sharp peaks at the opening angles determined by the zeros of
the Bessel functions $J_{m}(\lambda _0\rho _1)$. At the angles
away from these peaks the radiation intensity is relatively
suppressed by the factor $\gamma _{1\parallel }$. In the weak
undulator regime the radiation is dominated by the fundamental
harmonic and the leading contribution into the intensity comes
from second term on the right of Eq. (\ref{Cm+p}). In the case of
a particle moving in a homogeneous medium this term vanishes and
it is necessary to take the subleading terms. The corresponding
numerical results are plotted in Figures \ref{fig3} and
\ref{fig4}. These results show that the presence of the cylinder
provides a possibility for an essential enhancement of the
radiated power as compared to the radiation in a homogeneous
medium.

\section*{Acknowledgement}

The authors are grateful to Professor A.R. Mkrtchyan for general
encouragement and to Professor L.Sh. Grigoryan, S.R. Arzumanyan,
H.F. Khachatryan for stimulating discussions. A.A.S. acknowledges
the hospitality of the Abdus Salam International Centre for
Theoretical Physics, Trieste, Italy. The work has been supported
by Grant No.~1361 from Ministry of Education and Science of the
Republic of Armenia.

\end{document}